\documentclass[twocolumn, tighten, times]{aastex631}

\shorttitle{ICRF Photometric Variability}
\shortauthors{Secrest}

\begin{document}

\title{Optical-Radio Position Offsets are Inversely Correlated with AGN Photometric Variability}

\correspondingauthor{Nathan Secrest}
\email{nathan.j.secrest.civ@us.navy.mil}

\author[0000-0002-4902-8077]{Nathan J.\ Secrest}
\affiliation{U.S.\ Naval Observatory,
3450 Massachusetts Ave., NW,
Washington, DC 20392, USA}

\begin{abstract}
Using photometric variability information from the new Gaia~DR3 release, I show for the first time that photometric variability is inversely correlated with the prevalence of optical-radio position offsets in the active galactic nuclei (AGNs) that comprise the International Celestial Reference Frame (ICRF). While the overall prevalence of statistically significant optical-radio position offsets is 11\%, objects with the largest fractional variabilities exhibit an offset prevalence of only $\sim2\%$. These highly variable objects have redder optical color and steeper optical spectral indices indicative of blazars, in which the optical and radio emission is dominated by a line-of-sight jet, and indeed nearly $\sim100\%$ of the most variable objects have $\gamma$-ray emission detected by Fermi~LAT. This result is consistent with selection on variability preferentially picking jets pointed closest to the line-of-sight, where the projected optical-radio position offsets are minimized and jet emission is maximally boosted in the observed frame. While only $\sim9\%$ of ICRF objects exhibit such large photometric variability, these results suggest that taking source variability into account may provide a means of optimally weighting the optical-radio celestial reference frame link.
\end{abstract}

\section{Introduction}
Development and refinement of the International Celestial Reference Frame \citep[ICRF;][]{1998AJ....116..516M}, which realizes the International Celestial Reference System \citep[ICRS;][]{1995A&A...303..604A} via the positions of distant radio active galactic nuclei (AGNs) measured with very long baseline interferometry (VLBI), is an ongoing collaborative activity of several international research institutions, including the U.S.\ Naval Observatory. Despite their particular importance for celestial navigation, astrometry, and geodesy, surprisingly little is known about the physical characteristics of the AGNs that comprise the ICRF, such as their AGN spectral types, black hole masses, accretion rates, host galaxies, and (in many cases) redshifts \citep{2022ApJS..260...33S}. One outstanding issue that complicates efforts to produce a unified, wavelength-independent reference frame\footnote{\url{https://www.iau.org/static/science/scientific_bodies/working_groups/329/charter_icrf-multiwaveband-wg.pdf}} is the presence of statistically significant optical-radio position offsets on milliarcsecond scales (1~mas~$\sim8$~pc for typical moderate-redshift sources), which became especially conspicuous with the advent of the European Space Agency's Gaia astrometric space observatory \citep{2017ApJ...835L..30M}. A major source of these offsets may be variations in the apparent position of the radio core due to synchrotron self-absorption \citep[e.g.,][]{2019MNRAS.485.1822P}, and indeed several works have demonstrated correlations between the size and position angle of optical-radio offsets and parsec-scale radio source extent \citep{2017A&A...598L...1K}. Independent of variations in the apparent structure of the jet, AGN photometric variability may induce apparent motion \citep[e.g.,][]{2016ApJS..224...19M} due to changing relative contributions of AGN versus host galaxy emission, or the presence of secondary AGNs in a dual system. On the other hand, AGNs with very large variability amplitudes may exhibit fewer optical-radio offsets, as large variability is seen in blazars, AGNs for which the jet is closely aligned with the line-of-sight (LOS). Finally, photometric variability has been suggested to be a potential indicator of changes in the radio source structure \citep[e.g.,][]{2018A&A...611A..52T}.

In this Letter, I determine for the first time the relationship between photometric variability and optical-radio offset prevalence, using Gaia~DR3 data of ICRF3 \citep{2020A&A...644A.159C} objects in the Gaia Celestial Reference Frame \citep[Gaia-CRF3;][]{2022arXiv220412574G}. Section~\ref{sec: methodology} describes the data used in this study, the variability metric employed, and the methodology for producing an astrometrically clean sample of ICRF objects with Gaia counterparts. Results are discussed in Section~\ref{sec: results}, with the main conclusions in Section~\ref{sec: conclusions}.

\section{Methodology} \label{sec: methodology}
I cross-matched the positions of the 4536 sources in ICRF3~S/X with Gaia-CRF3 to within 100~mas, obtaining 3142 objects, as in \citet{2022arXiv220412574G}. Joining these on the DR3 QSO candidates table, 2426 have a measurement of the fractional variability in the Gaia $G$ band (\texttt{fractional\_variability\_g}; hereafter $F_\mathrm{var}$), which I use for this study. This is defined as:

\begin{equation} \label{eq: Fvar}
F_\mathrm{var} = \frac{\sqrt{\mathrm{MAD}^2(F) - \langle\sigma_F^2\rangle}}{\mathrm{median}(F)}
\end{equation}

\noindent where $F$ is the flux in the Gaia $G$ band, $\langle\sigma_F^2\rangle$ is the mean of the flux variances, and $\mathrm{MAD}(F)$ is the median~absolute~deviation, a robust estimator of the standard deviation \citep{2022arXiv220706849C}.\footnote{\url{https://gea.esac.esa.int/archive/documentation/GDR3/Gaia_archive/chap_datamodel/sec_dm_variability_tables/ssec_dm_vari_agn.html}} In other words, $F_\mathrm{var}$ is a measure of excess variability not attributable to the formal photometric uncertainties. Note, however, that the factor of $\sim1.5$ used to convert $\mathrm{MAD}$ to the standard deviation is missing from Equation~\ref{eq: Fvar}, which causes the excess variability to be underestimated in an absolute sense. As noted in \citet{2022arXiv220706849C}, this is partially mitigated by Gaia photometric uncertainties being underestimated to a degree, but in the present work only relative differences in $F_\mathrm{var}$ matter, not the absolute values.

In order to determine the prevalence of optical-radio offsets, I calculated the normalized astrometric offsets $X$, which take into account the full positional covariance, following Equation~4 in \citet{2016A&A...595A...5M}. For positions with no intrinsic offset and accurate uncertainties, $X$ should follow a Rayleigh distribution with $\sigma=1$, denoted as ${\cal R}(1)$. Then, the fraction of objects above some value of $X$ expected by chance is calculated using the survival function, $1-\mathrm{CDF}_{\mathcal{R}(1)}(X)$. For $X>5$, this is $3.7\times10^{-6}$, so there should be no false positives in a catalog of size 2426. In fact, however, the prevalence of ICRF3~S/X objects in Gaia-CRF3 with $X > 5$ is 13\%.

To avoid confounding factors that may induce optical-radio offsets, such as spurious astrometry due to source extent or source multiplicity (e.g., dual AGNs), I applied several quality cuts. First, I required that the BP/RP excess factor (\texttt{phot\_bp\_rp\_excess\_factor}) be less than 2, which helps eliminate extended sources, such as those at lower redshifts \citep{2022A&A...660A..16S, 2022ApJ...933...28M}. Second, I required that the significance of any astrometric excess noise (\texttt{astrometric\_excess\_noise\_sig}) be less than 2. Significant astrometric excess noise has been used to select close dual quasars \citep[e.g.,][]{2021NatAs...5..569S}, which are a potential source of optical-radio offsets. I further removed sources exhibiting multiple peaks in the Gaia scanning windows by requiring that \texttt{ipd\_frac\_odd\_win} and \texttt{ipd\_frac\_multi\_peak} be zero. Further details of these parameters are given in the Gaia~DR3 data model documentation.\footnote{\url{https://gea.esac.esa.int/archive/documentation/GDR3/Gaia_archive/chap_datamodel/sec_dm_main_source_catalogue/ssec_dm_gaia_source.html}} 

Finally, apparently significant parallaxes and proper motions in bona fide quasars are an indication of multiplicity, such as dual AGNs or gravitational lenses \citep{2022A&A...660A..16S, 2022ApJ...933...28M}, so should be removed. Note that calculation of the uncertainty-normalized parallaxes and proper motions first requires correction for the zeropoint offsets and uncertainty scaling factors. This is done iteratively, adjusting the offset and scaling factors using objects with normalized values consistent with expectations for the size of the sample, until convergence to the normal distribution $\mathcal{N}(0,1)$. For the 1953 ICRF3~S/X objects in Gaia-CRF3 that meet the source extent/multiplicity cuts, not one should have absolute normalized values exceeding $3.5$. The results of this process are given in the Appendix in Table~\ref{tab: offsets}. The offsets and parallax uncertainty scaling factor are generally consistent with previous literature estimates using large quasar samples \citep{2022A&A...660A..16S, 2022ApJ...933...28M}, but the proper motions scaling factors, about a factor of 1.14, are significantly larger than the literature estimates of about 1.06 \citep[see also][]{2021A&A...649A...9G}. The value of 1.06 from the literature was obtained using much larger and heterogeneous samples of AGNs and quasars, namely Gaia-CRF3, LQAC-5 \citep{2019A&A...624A.145S}, and mid-IR AGNs \citep{2015ApJS..221...12S} \citep[][respectively]{2021A&A...649A...9G,2022A&A...660A..16S,2022ApJ...933...28M}, suggesting that 1.06 is a more accurate estimate of the degree to which Gaia EDR3/DR3 astrometric uncertainties are underestimated due to data processing issues. The larger scaling factors found here may therefore reflect genuine astrophysical variance particular to the radio-bright objects that comprise the ICRF.

With these considerations, I corrected for the parallax offset, which is significant, and applied a 1.06 correction factor to the parallax and proper motion uncertainties. I removed three objects that have absolute normalized parallaxes greater than 3.5, which are recorded in Table~\ref{tab: icrf_para}. The corresponding cut for $\mathcal{R}(1)$-distributed normalized proper motions is $\sim3.9$ for 1950 objects, which yields 12 that are significant, listed in Table~\ref{tab: icrf_pm}. Notably, one of the ICRF3 defining sources is counted among these 12 objects, with a normalized proper motion of $\chi=4.3$, which has an expected frequency in $\mathcal{R}(1)$-distributed values of $9.7\times10^{-4}$. Removing these, there are 1938 objects in the final sample.

Note that these results suggest that the position uncertainties $\sigma_{\alpha^*}$ and $\sigma_\delta$ may also be underestimated by $\sim6\%$, which would cause the prevalence of optical-radio position offsets to be overestimated. Even scaling the position uncertainties up by 6\%, however, the prevalence of optical-radio position offsets in the final sample is 11\%. To remain conservative, I assume that the Gaia EDR3/DR3 position uncertainties are indeed underestimated, and adopt 11\% as the fiducial overall prevalence of optical-radio offsets.

\section{Results} \label{sec: results}
\begin{figure*}
  \centering
   \includegraphics[width=\textwidth]{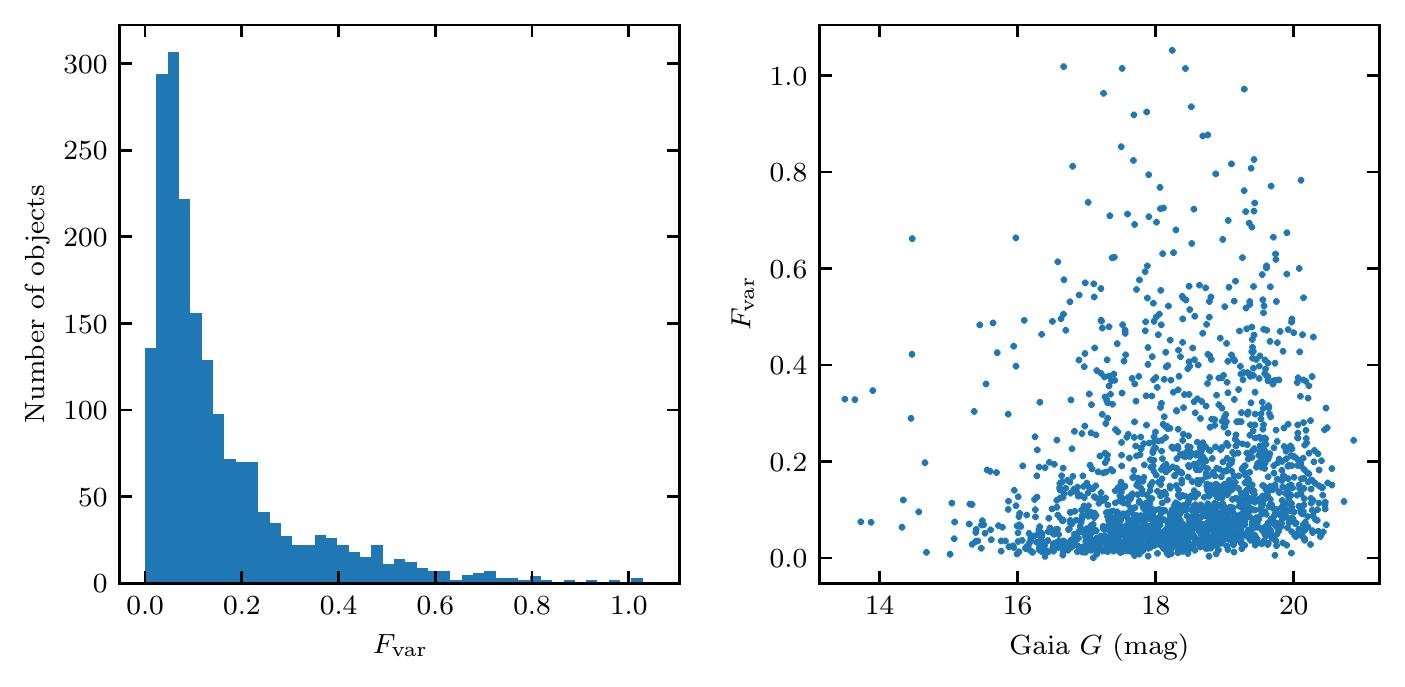}
   \caption{Left: distribution of fractional excess variability in the Gaia~$G$ band for the 1938 ICRF3~S/X objects in Gaia-CRF3 meeting the astrometric quality cuts employed here. Right: fractional excess variability as a function of $G$, indicating a very weak dependence.}
   \label{fig:fvar_g}
\end{figure*}

The left panel of Figure~\ref{fig:fvar_g}, shows the distribution of $F_\mathrm{var}$ in the 1938 ICRF3~S/X objects in Gaia-CRF3 that meet the astrometric quality cuts imposed in Section~\ref{sec: methodology}. Reiterating that $F_\mathrm{var}$ is the fractional photometric variability in excess of expectations from the uncertainties, all of the ICRF objects exhibit some level of variability, with a mean of 16\% (0.17~mag). Importantly, as shown in the right panel of Figure~\ref{fig:fvar_g}, the minimum detectable $F_\mathrm{var}$ is nearly unbiased with respect to $G$ magnitude, mitigating differences in source brightness as a confounding factor in this analysis. 

\begin{figure}
  \centering
   \includegraphics[width=\columnwidth]{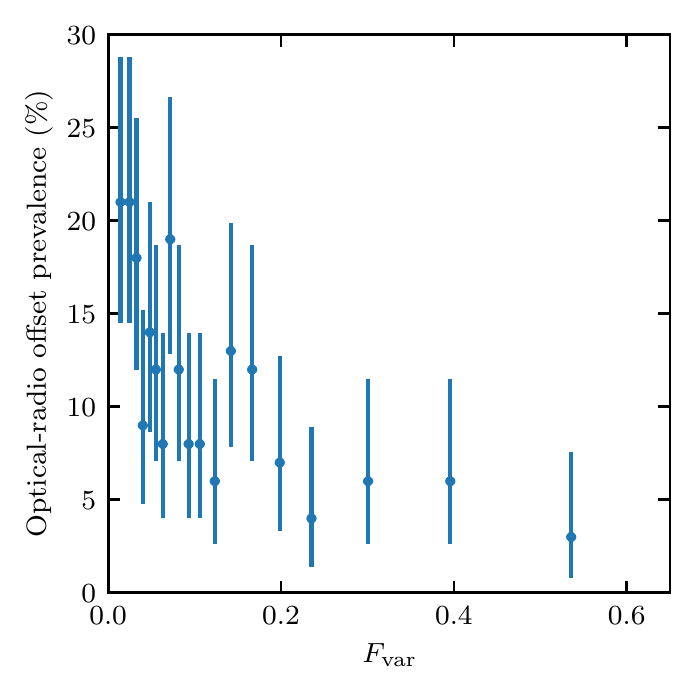}
   \caption{Percentage of objects with significant optical-radio position offsets ($X>5$), in $F_\mathrm{var}$ bins of 100 objects. Error bars denote the 90\% CI.}
   \label{fig:X_vs_Fvar_binned}
\end{figure}

It is immediately clear that the prevalence of optical-radio position offsets is inversely correlated with $F_\mathrm{var}$, as shown in Figure~\ref{fig:X_vs_Fvar_binned} where the sample is binned with respect to $F_\mathrm{var}$, requiring 100 objects per bin. I used binomial statistics \citep[e.g.,][]{1986ApJ...303..336G} to determine the 90\% confidence interval (CI) of the population mean prevalence in each bin and assess the significance of differences in prevalence as a function of $F_\mathrm{var}$. A Kendall's $\tau$ test for the unbinned data gives a $p$-value of $10^{-6}$, firmly rejecting the null hypothesis that there is no relationship between position offsets and fractional variability. Objects with very little or no fractional photometric variability exhibit a position offset prevalence of $\sim20\%$, dropping to a few percent for objects above $F_\mathrm{var}\gtrsim0.4$.

\begin{figure}
  \centering
   \includegraphics[width=\columnwidth]{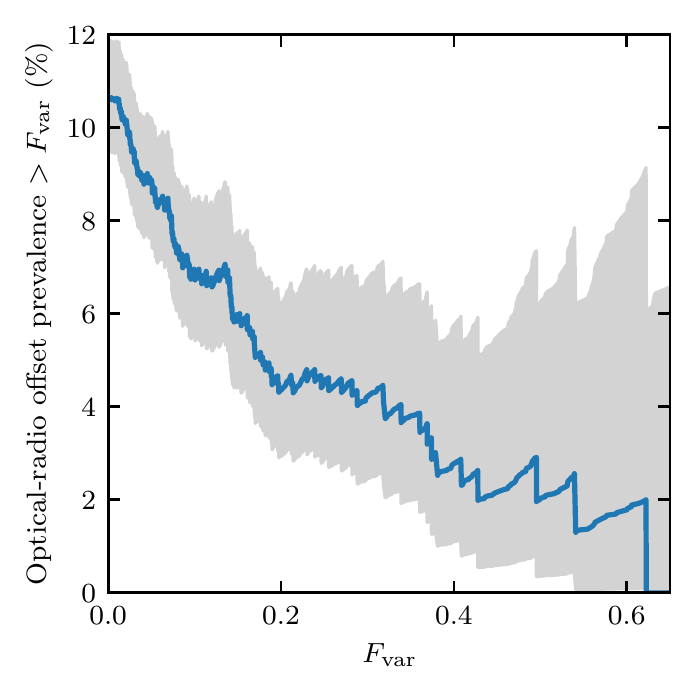}
   \caption{Percentage of objects above the fractional variability $F_\mathrm{var}$ with significant optical-radio position offsets ($X>5$). The gray shaded region denotes the 90\% CI.}
   \label{fig:X_vs_Fvar}
\end{figure}

For practical applications, it is preferable to know the rate of offsets for all objects above some $F_\mathrm{var}$ threshold, so I sorted the ICRF objects by increasing $F_\mathrm{var}$ and determined the fraction of objects above each value of $F_\mathrm{var}$ that have offsets. As before, there is a large, significant, and possibly monotonic drop in the prevalence of optical-radio position offsets with increasing photometric variability (Figure~\ref{fig:X_vs_Fvar}). This drop reaches a statistical minimum of $(2.0^{+3.1}_{-1.4})\%$ for objects with $F_\mathrm{var} \gtrsim 0.4$, where the uncertainty bounds contain the 90\% CI. Beyond this value of $F_\mathrm{var}$, the sample prevalence of optical-radio offsets continues to drop, but the uncertainties of the true population mean grow due to small number statistics. Splitting the sample into three fractional variability categories, 75\% have low variability ($F_\mathrm{var} < 0.2$), 9\% have high variability ($F_\mathrm{var} > 0.4$), and 16\% are intermediate.

\begin{figure}
  \centering
   \includegraphics[width=\columnwidth]{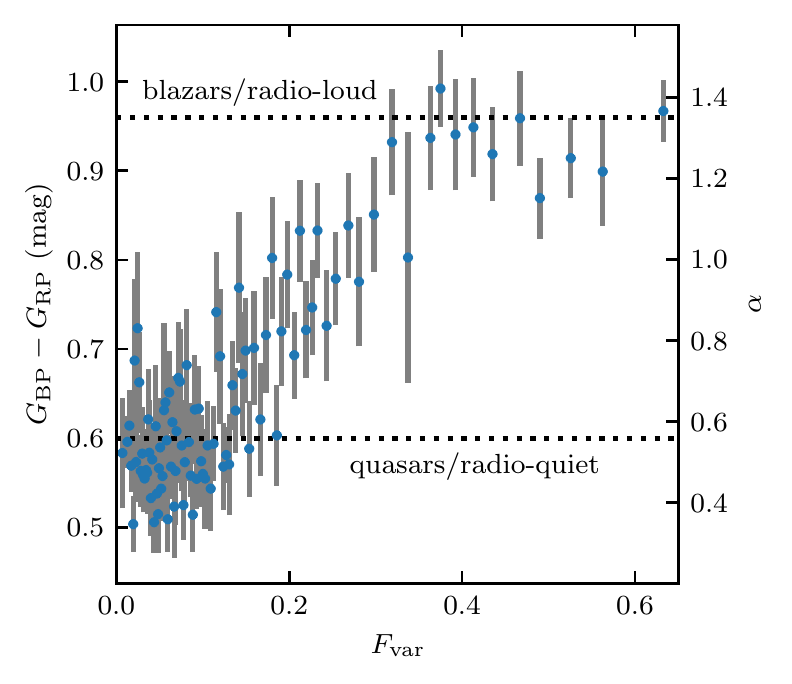}
   \caption{Gaia $G_\mathrm{BP} - G_\mathrm{RP}$ color/spectral index $\alpha$ ($S_\nu \propto \nu^{-\alpha}$) as a function of fractional variability. ICRF sources become progressively redder with increasing variability, reaching a plateau at $F_\mathrm{var}\sim0.4$, where they exhibit $G_\mathrm{BP} - G_\mathrm{RP}$ color consistent with blazars (dotted line, showing the median color for all $L_\gamma = 10^{46-47}$~erg~s$^{-1}$ blazars, using the templates from \citealt{2017MNRAS.469..255G}). This sequence is consistent with the trend of $\alpha$ versus radio loudness from \citet{2002AJ....124.2364I}. Gaia magnitudes are in the Vega system.}
   \label{fig:bprp_vs_Fvar}
\end{figure}

Concomitant with the drop in optical-radio offset prevalence with fractional variability, ICRF objects exhibit progressively redder Gaia $G_\mathrm{BP} - G_\mathrm{RP}$ color, as shown in Figure~\ref{fig:bprp_vs_Fvar}. Above $F_\mathrm{var} > 0.4$, the $G_\mathrm{BP} - G_\mathrm{RP}$ color plateaus at a median value of 0.94~mag (Vega). Using the empirical blazar spectral energy distribution (SED) parameterizations of \cite{2017MNRAS.469..255G} and calculating the synthetic $G_\mathrm{BP} - G_\mathrm{RP}$ color for the full range of blazar luminosities, this median color is expected for blazars with a $\gamma$-ray luminosity between $10^{46} - 10^{47}$~erg~s$^{-1}$, nearly independent of redshift (upper dotted line in Figure~\ref{fig:bprp_vs_Fvar}). The objects at small values of $F_\mathrm{var}$ ($\lesssim0.2$) have $G_\mathrm{BP} - G_\mathrm{RP} \sim 0.6$, which is the typical color of moderate-redshift quasars selected from the the Sloan~Digital~Sky~Survey \citep[SDSS;][]{2017AJ....154...28B} and  Baryon~Oscillation~Spectroscopic~Survey \citep[BOSS;][]{2016AJ....151...44D} \texttt{specObj} table for DR17.\footnote{\url{https://www.sdss.org/dr17/spectro/spectro_access}} This result is consistent with studies of SDSS objects with radio counterparts, which have shown that the optical spectral index $\alpha$ is generally larger/steeper for radio-loud objects, compared to radio-quiet \citep[e.g., Figure~17 in][]{2002AJ....124.2364I}. 

\begin{figure}
  \centering
   \includegraphics[width=\columnwidth]{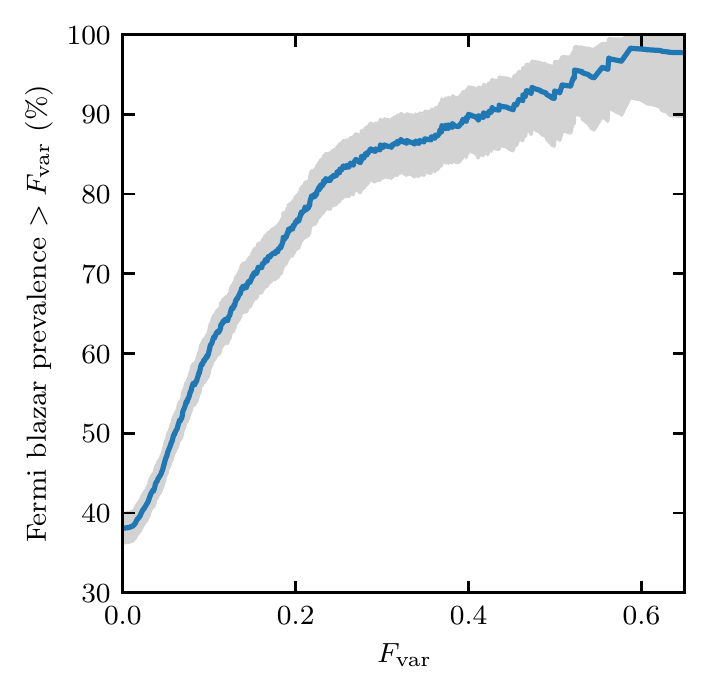}
   \caption{Percentage of objects above the variability threshold identified as a Fermi blazar. The gray shaded region denotes the 90\% confidence interval, determined using binomial statistics.}
   \label{fig:blazar_vs_Fvar}
\end{figure}

I checked this result by matching to the fourth Fermi Large Area Telescope (LAT) catalog, DR3 \cite[4FGL-DR3;][]{2022ApJS..260...53A}, which contains 6,659 $\gamma$-ray sources distributed across the full sky. There are 4,502 objects in 4FGL-DR3 with an identified counterpart, 738 of which have a match to the 1938 ICRF sources within $1\arcsec$. I find no evidence for a difference in the distribution of $G$ magnitudes between matches and non-matches, with a K-S test giving $p=0.75$, indicating that the matches are not biased against fainter sources. In Figure~\ref{fig:blazar_vs_Fvar}, I show the percentage of objects above a given fractional variability threshold identified as a blazar, based on having a $\gamma$-ray counterpart in 4FGL-DR3. While 38\% of all 1938 objects are identified as a Fermi blazar, this percentage steadily grows to $\sim100\%$ at the highest variability thresholds. By $F_\mathrm{var} > 0.4$, at which point the incidence of optical-radio offsets is $\sim2\%$ and the median optical color is consistent with expectations for blazars, 90\% of ICRF objects are identified Fermi blazars.

Finally, note that the relationship between optical-radio offset prevalence and blazar classification is sensitive to how an object is classified as a blazar. Of the 1938 objects studied here, 1565 are flagged as a blazar in the Gaia-CRF3 cross-match table \texttt{gaia\_crf3\_xm}, which has objects listed in the BZCAT5 \citep{2015Ap&SS.357...75M}, the 2WHSP \citep{2017A&A...598A..17C}, and the ALMA calibrator catalog \citep{2019MNRAS.485.1188B}, the latter of which are considered to be mostly blazars. Despite 81\% of objects being a member of at least one of these catalogs, the rate of optical-radio offsets is the same as the full sample, 11\%. The resolution to this apparent contradiction is likely in how blazars are being defined. BZCAT blazars are selected via a wide range of criteria, from optical spectral properties to radio spectral index \citep{2009A&A...495..691M}. 2WHSP blazars were selected on their radio through X-ray SEDs to determine the peak frequency of synchrotron emission. ALMA calibrator catalog objects are compact radio sources, most of which are found either in BZCAT or were found to have flat radio spectra \citep{2018MNRAS.478.1512B, 2019MNRAS.485.1188B}. For all three catalogs, variability was not included in the respective selection criteria, and indeed of the 1565 flagged blazars, 1169 have low variability ($F_\mathrm{var} < 0.2$), while only 151 have high variability ($F_\mathrm{var} > 0.4$). By comparison, of the 738 Fermi blazars in the sample, 162 have $F_\mathrm{var} > 0.4$, while 369 have $F_\mathrm{var} < 0.2$. $\gamma$-ray selected blazars are thus $\sim9$ times more likely to show large fractional variability than blazars selected through varied methods, consistent with having large Lorentz factors and highly beamed emission. In this picture, small variations in the bulk Doppler factor due to, e.g., turbulence and instabilities in the jet lead to dramatic variations in the observed brightness of the blazar. As the Doppler factor is maximized when the jet is co-aligned with the LOS, large fractional variability is the hallmark of a jet pointed directly at the observer.

\section{Conclusions} \label{sec: conclusions}
I have quantified for the first time the relationship between optical photometric variability and the incidence of optical-radio position offsets in the ICRF, using epoch photometry from the new Gaia~DR3 catalog. Comparing the Gaia $G$-band fractional variability with the prevalence of optical-radio offsets in a sample of 1938 astrometrically clean sources shared between ICRF3~S/X and Gaia-CRF3, my primary results are as follows:

\begin{enumerate}
\item The prevalence of statistically significant optical-radio position offsets drops rapidly and possibly monotonically with increasing fractional photometric variability. At a variability level of $>40\%$, the prevalence of optical-radio offsets is $\sim2\%$, compared to $\sim11\%$ in the overall population.
\item The optical color/spectral index is strongly dependent on photometric variability, the most variable objects being typically $\sim0.3$ mag redder, with a spectral index ($S_\nu \propto \nu^{-\alpha}$) of $\alpha\sim1.3$, compared to $\alpha\sim0.5$ for the least variable objects. This is consistent with the most variable objects being radio-loud blazars, while the least variable objects have color/spectral index consistent with radio-quiet quasars.
\item Concordantly, the rate of objects with $\gamma$-ray-selected blazars from the all-sky Fermi~LAT catalog is nearly $\sim100\%$ for the most variable objects in the ICRF, monotonically increasing from a frequency of 38\% for the overall ICRF/Gaia catalog. The large Lorentz factors implied by $\gamma$-ray brightness can induce large fractional variability in the observed source brightness for small LOS jet angles.
\end{enumerate}

These results are consistent with a simple physical model of the relationship between the radio VLBI and optical Gaia source positions. For the 75\% of ICRF objects exhibiting low fractional photometric variability, the optical Gaia counterpart is the compact accretion disk and broad line region seen in the optical spectra of typical quasars. The radio jet is not aligned with our LOS, so the apparent core position measured in VLBI sessions may therefore be strongly subject to ``core shift'' caused by synchrotron self-absorption. This phenomenon occurs on milliarcsecond scales \citep[e.g.,][]{2011A&A...532A..38S}, the same as the typical scale of the optical-radio position offsets. The 9\% of ICRF objects that exhibit large variability, on the other hand, are almost entirely bona fide blazars, where the accretion disk, radio core, and extended jet are closely aligned with the LOS, minimizing the projected separation between the optical and radio source positions. The optical position is also more likely to be dominated by boosted jet emission, further reducing disagreement with the radio position.
Note that this physical picture additionally predicts that low-variability objects not currently exhibiting a significant optical-radio offset are more likely to develop an offset at later epochs, while the high-variability objects, barring rapid jet precession out of the LOS, are unlikely to develop optical-radio offsets in the future.

While accounting for photometric variability allows for the prevalence of optical radio offsets to be reduced from $\sim11\%$ to $\sim2\%$, it is important to reiterate that the number of highly variable sources is small, 180 in ICRF3~S/X, likely precluding use of these objects alone in the construction of a multi-wavelength CRF. There are two potential applications of these results that may nonetheless significantly improve the radio/optical CRF alignment. The first is factoring variability into the weighting of sources for the frame alignment, tying the frames more strongly between the highly variable sources and down-weighting the lower variability sources appropriately to account for their tendency to exhibit or develop significant position offsets. The second is to draw preferentially from highly variable sources in the construction of the next ICRF. The 4536 objects in ICRF3~S/X are a small subset of the $\sim$~20,000 VLBI sources currently known, so observing priority could be given to those prospective sources that are, a priori, less likely to have optical-radio position offsets.

\begin{acknowledgements}
I thank the anonymous referee and statistician for important comments that significantly improved this work, as well as Valeri Makarov for helpful discussions. This research made use of Astropy,\footnote{http://www.astropy.org} a community-developed core Python package for Astronomy \citep{2013A&A...558A..33A, 2018AJ....156..123A}.
\end{acknowledgements}

\facilities{Gaia}

\software{Astropy \citep{2013A&A...558A..33A, 2018AJ....156..123A}, dustmaps \citep{2018JOSS....3..695G}, \textsc{topcat} \citep{2005ASPC..347...29T}}

\bibliography{icrf_var_nsecrest}
\clearpage
\appendix

\begin{deluxetable}{ccrc} \label{tab: offsets}
\tablehead{\colhead{parameter} & \colhead{unit} & \colhead{offset} & \colhead{scaling}}
\startdata
$\varpi$ & $\mu$as & $-24$\,(3) & 1.061\,(0.018) \\
$\mu_{\alpha^*}$ & $\mu$as yr$^{-1}$ & $+1$\,(4) & 1.134\,(0.022) \\
$\mu_\delta$ & $\mu$as yr$^{-1}$ & $-1$\,(3) & 1.148\,(0.021)
\enddata
\caption{Parallax and proper motion offsets and scaling factors for Gaia-CRF3 counterparts of ICRF3~S/X objects meeting the criteria given in Section~\ref{sec: methodology} to exclude source multiplicity. One-sigma uncertainties, estimated using $10^4$ random samples with replacement, are given in parentheses. $\alpha =$~right ascension and $\delta = $ declination, with $\alpha^*$ indicating multiplication by $\cos(\delta)$ when expressing differences of right ascension in degrees of arc.}
\end{deluxetable}

\begin{deluxetable*}{cccccccr} \label{tab: icrf_para}
\tablehead{\colhead{IERS} & \colhead{defining} & \colhead{$\alpha$} & \colhead{$\delta$} & \colhead{source\_id} & \colhead{$\varpi$} & \colhead{$\varpi/\sigma_\varpi$} & \colhead{$X$}\\ \colhead{ } & \colhead{ } & \colhead{$\mathrm{{}^{\circ}}$} & \colhead{$\mathrm{{}^{\circ}}$} & \colhead{ } & \colhead{$\mathrm{mas}$} & \colhead{ } & \colhead{ }}
\startdata
0829$+$046 &  & 127.95365 & $+$4.49419 & 3092416108553785600 & $-$0.200 & $-$3.5 & 1.8 \\
0918$-$363 &  & 140.10917 & $-$36.52987 & 5431288820729101056 & $+$0.568 & $+$3.7 & 3.9 \\
1807$+$698 &  & 271.71117 & $+$69.82447 & 2260127244173131520 & $-$0.081 & $-$3.6 & 14.7
\enddata
\caption{Gaia-CRF3 counterparts of ICRF3~S/X objects with absolute uncertainty-normalized parallaxes $\varpi/\sigma_\varpi > 3.5$, potentially indicating complex or problematic Gaia EDR3/DR3 astrometry. Note that one of the objects also has a significant optical-radio position offset ($X>5$)}
\end{deluxetable*}

\begin{deluxetable*}{ccccccccccr} \label{tab: icrf_pm}
\tablehead{\colhead{IERS} & \colhead{defining} & \colhead{$\alpha$} & \colhead{$\delta$} & \colhead{source\_id} & \colhead{$\mu_{\alpha^*}$} & \colhead{$\sigma_{\mu_{\alpha^*}}$} & \colhead{$\mu_\delta$} & \colhead{$\sigma_{\mu_\delta}$} & \colhead{$\chi$} & \colhead{$X$}\\ \colhead{ } & \colhead{ } & \colhead{$\mathrm{{}^{\circ}}$} & \colhead{$\mathrm{{}^{\circ}}$} & \colhead{ } & \colhead{$\mathrm{mas\,yr^{-1}}$} & \colhead{$\mathrm{mas\,yr^{-1}}$} & \colhead{$\mathrm{mas\,yr^{-1}}$} & \colhead{$\mathrm{mas\,yr^{-1}}$} & \colhead{ } & \colhead{ }}
\startdata
0110$+$495 & D & 18.36253 & $+$49.80668 & 403495100669336960 & $+$0.233 & 0.073 & $-$0.248 & 0.078 & 4.3 & 5.4 \\
0922$+$316 &  & 141.43188 & $+$31.45300 & 700509750093550208 & $+$1.698 & 0.444 & $-$0.447 & 0.373 & 4.7 & 12.8 \\
1048$+$347 &  & 162.74218 & $+$34.50304 & 738378339303790720 & $+$2.576 & 0.752 & $-$2.696 & 0.892 & 3.9 & 3.4 \\
1302$-$102 &  & 196.38756 & $-$10.55540 & 3622979843899961472 & $-$0.040 & 0.042 & $+$0.132 & 0.030 & 4.6 & 6.2 \\
1335$+$552 &  & 204.45684 & $+$55.01725 & 1562088440305113984 & $-$0.269 & 0.087 & $+$0.220 & 0.103 & 4.1 & 3.6 \\
1641$+$399 &  & 250.74504 & $+$39.81028 & 1355746700891999360 & $-$0.478 & 0.127 & $-$0.360 & 0.152 & 4.2 & 49.5 \\
1657$+$022 &  & 254.93749 & $+$2.21862 & 4385031626327045120 & $-$0.609 & 0.168 & $-$0.463 & 0.105 & 4.7 & 1.4 \\
1741$+$279 &  & 265.98516 & $+$27.88065 & 4594710490905858048 & $+$0.128 & 0.060 & $+$0.268 & 0.076 & 4.0 & 3.5 \\
1920$-$211 &  & 290.88412 & $-$21.07593 & 6772838283893505152 & $-$0.176 & 0.160 & $-$0.505 & 0.134 & 3.9 & 4.1 \\
2201$+$315 &  & 330.81240 & $+$31.76063 & 1899428400235300608 & $+$0.090 & 0.022 & $+$0.037 & 0.031 & 4.2 & 4.6 \\
2312$-$319 &  & 348.70209 & $-$31.64431 & 6556811560722517632 & $-$0.431 & 0.125 & $+$0.150 & 0.144 & 4.0 & 2.0 \\
2346$+$052 &  & 357.33771 & $+$5.57774 & 2744495750095600256 & $-$0.171 & 0.202 & $+$0.418 & 0.105 & 4.6 & 2.6
\enddata
\caption{Gaia-CRF3 counterparts of ICRF3~S/X objects with uncertainty-normalized proper motions $\chi > 4$, potentially indicating significant proper motion. Note that 4 out of 12 objects also have significant optical-radio position offsets ($X>5$).}
\end{deluxetable*}

\end{document}